\documentstyle[12pt,epsfig]{article}
\newcommand{\bc}{\begin{center}}
\newcommand{\ec}{\end{center}}
\newcommand{\bt}{\begin{tabular}}
\newcommand{\et}{\end{tabular}}
\newcommand{\be}{\begin{equation}}
\newcommand{\ee}{\end{equation}}
\newcommand{\bd}{\begin{displaymath}}
\newcommand{\ed}{\end{displaymath}}
\newcommand{\ba}{\begin{array}}
\newcommand{\ea}{\end{array}}

\def\fun#1#2{\lower3.6pt\vbox{\baselineskip0pt\lineskip.9pt
\ialign{$\mathsurround=0pt#1\hfil##\hfil$\crcr#2\crcr\sim\crcr}}}

\def\baselinestretch{1.2}
\begin{document}

{

\renewcommand\baselinestretch{1}

\bc

GRIBOV THEORY OF NUCLEAR INTERACTIONS AND PARTICLE DENSITIES
AT FUTURE HEAVY-ION COLLIDERS .\\
\vspace{0.4cm}

A. Capella$^{a)}$, A. Kaidalov$^{a),b)}$ and J. Tran Thanh Van$^a)$\\

a)~ Laboratoire de Physique Th\'eorique,\\
Unit\'e Mixte de Recherche  (CNRS) UMR 8627 \\
Universit\'e de Paris XI, B\^atiment 210, F-91405 Orsay Cedex, France\\ 
b)~ ITEP, B.Cheremushkinskaya 25, 117526 Moscow, Russia

\ec

\vspace{1cm}

\begin{abstract}

 Gribov approach to high--energy interactions of hadrons and nuclei is reviewed
and applied to calculation of particle production in heavy--ions collisions.
It is pointed out that the AGK (Abramovsky, Gribov, Kancheli) cutting rules is
a po\-wer\-ful tool to investigate particle spectra in these processes. It
leads, in the Glauber approximation, to a simple formula for the density of hadrons
produced in the central rapidity region in nucleus--nucleus interactions.
 An estimate of this density for RHIC and LHC is presented and compared with
 results of Monte--Carlo calculations. It is shown that the Glauber
 approximation substantially overestimate  particle densities
 compared to the results of the complete Gribov theory. This is due to extra
shadowing in the system, related to large mass diffraction which leads to a strong
decrease of particle densities at mid rapidities. Our method of calculation
of these effects has been applied to the problem of shadowing
 of nuclear structure functions and a good agreement with experimental data has been obtained. 
\end{abstract}
\vskip 1cm
\begin{center}
Contributed paper to the\\ {\it Gribov Memorial Volume}\\ of Acta Physica Hungarica
(Heavy Ion Physics).
\end{center}

\vskip 1 truecm
\noindent LPT ORSAY 99-15 \par
\noindent March 1999     
 \vskip 2 truecm
\newpage
\section{Introduction}

The reggeon approach to high--energy interactions is undoubtly 
an important ingredient of modern theory. V.N. Gribov has made very essential
contributions to the development of this approach. He has introduced a leading pole in
the complex angular momentum plane \cite{Gribov1}, which determines asymptotics
of diffractive processes (nowadays called the Pomeron), investigated
the main properties of Regge--poles \cite{Gribov2} and Regge--cuts \cite{Gribov3} and
developed the reggeon diagram technique \cite{Gribov4}.

An important contribution to the theory of multiparticle production has been made
by V.N. Gribov together with V.A. Abramovsky and O.V. Kancheli \cite{AGK} 
and is usually referred to as the AGK--cutting rules. These rules are widely used
for the construction of QCD--based models for high--energy interactions
(for reviews see \cite{KaidalovR,CapellaR}). In this paper we will extensively
 use AGK--cutting
rules and Gribov reggeon diagram technique to calculate inclusive
particle spectra and particle densities in the central rapidity region for
heavy--ions collisions at future colliders RHIC and LHC. The values of particle
and energy densities are very important for the problem of creation of the
 quark--gluon plasma in heavy--ions collisions at these very high energies.

In the first part of the paper (Section 2) we will shortly review Gribov's
 approach to high--energy interactions of hadrons with nuclei \cite{Gribov5} 
and note the difference in the space--time picture of interactions at low and 
high energies.
We will also discuss a difference between the Glauber approximation and the general
Gribov theory.
The AGK--cutting rules will be applied to  hadron--nucleus and
nucleus--nucleus collisions and a simple formula for inclusive particle
spectra and densities of produced particles in the central rapidity region
valid in the Glauber approximation will be presented.

Applications of the Regge--Gribov theory to high--energy hadronic interactions are
considered in Section 3. The important role of shadowing effects in these
processes is emphasized. It is pointed out that understanding diffractive
processes is important for a selfconsistent description of high--energy
interactions of hadrons and nuclei. The results of this analysis are used in
the Section 4 to estimate particle
densities at RHIC and LHC . Limitations of the Glauber
approximation and its modification  at high energy according to the Gribov
theory will be discussed . It will be shown that these
modifications are related to large mass diffraction dissociation of hadrons
which leads to extra shadowing in the system.  These effects reduce
 particle densities in the central
rapidity region compared to the Glauber approximation predictions.
 This result is
valid for both soft and hard processes. To estimate
these shadowing effects we apply Gribov theory to the processes of deep
inelastic scattering (DIS) on nuclei and show how it is possible to calculate
nuclear shadowing effects in the region of small Bjorken--x, using information
on diffractive production in DIS. Our results agree with existing
experimental data and allow a safe extrapolation of shadowing effects to the
experimentally unmeasured region of very small x, relevant for nuclear
collisions at LHC. This model leads to a reduction of particles densities in PbPb collisions at LHC (RHIC)
by a factor of three (two), with respect to the results of the Glauber 
approximation. Our results are compared to predictions of different
Monte--Carlo calculations.

\section{Gribov theory of high-energy nuclear interactions}
 
 In classical papers of Gribov \cite{Gribov5} it was shown how to incorporate
 the Glauber model \cite{Glauber} for interactions of hadrons with nuclei
 into a general framework of relativistic quantum theory. Consider an
 amplitude of elastic scattering for high--energy hadron--nucleus 
 interactions. In the Glauber model it is described by the diagrams shown in
 Fig.~1, which looks like a successive rescatterings of initial hadron on
 nucleons of the nucleus. However, as was emphasized by Gribov \cite{Gribov5,Gribov6} 
the space--time picture of the interaction at high--energy 
 $E>m_h \mu R_A$ ($\mu$ is a characteristic hadronic scale $\sim 1~GeV$ and
 $R_A$ is the radius of the nucleus) is completely different from this 
 simple picture. It corresponds to coherent interactions of a fluctuation
 of the initial hadron, which is ``prepared" long before its interaction with
 the nucleus (Fig.~2). A very important result of Gribov \cite{Gribov5} is that
 nevertheless the elastic hA--amplitude can be written as a sum of the diagrams
 shown in Fig.~3, with elastic rescatterings (Fig.~3a) which give the
 same result as Glauber model, plus all possible diffractive excitations
 of the initial hadron. At not too high energies  $E_L\sim 10^2~GeV$ these
 terms lead to corrections to the Glauber approximation of $10-20\%$
 for the total $hA$ cross section \cite{Kaidalov1,Anisovich}.
 
 We will show below that at very high energies and for inclusive cross
 sections this modification of the Glauber approximation is very important.
 The difference between Glauber model and Gribov's theory is essential for
 understanding shadowing corrections for structure functions of nuclei 
 related to hard processes on nuclei and for many aspects of multiparticle
 production on nuclei \cite{Kaidalov}.
 
 An important consequence of the space--time structure of the diagrams of
 Fig.~2 for interactions of hadrons with nuclei is the theorem, based on
 AGK--cutting rules \cite{AGK}, that for inclusive cross sections
 all rescatterings cancel with each other and these cross sections are determined
  by the diagrams shown
 in Fig.~4 (impulse approximation). Note, however, that this result, valid asymptotically
 in the central rapidity region, only applies to the diagrams
 of the Glauber--type, i.e. when masses of intermediate states in Fig.~3 are
 limited and do not increase with energy. As a result, the inclusive cross
 section for the production of a hadron $a$ is expressed, for a given impact
parameter $b$, in terms of inclusive cross section for $hN$ interactions

\be
 E\frac{d^3\sigma^a_{hA}(b)}{d^3p}=T_A(b)E\frac{d^3\sigma^a_{hN}}{d^3p}
\label{eq1}
\ee

\noindent where $T_A(b)$ is the nuclear profile function $(\int d^2b~T_A(b)=A)$. After
integration over $b$ we get

\be
 E\frac{d^3\sigma^a_{hA}}{d^3p}=A~E\frac{d^3\sigma^a_{hN}}{d^3p}
\label{eq2}
\ee

The total and inelastic hA cross sections in the Glauber model can be easily
calculated and are given for heavy nuclei by well known expressions. For
example

\be
\sigma^{in}_{hA} = \int d^2b (1-exp(-\sigma^{in}_{hN}T_A(b))
\label{eq3}
\ee

The situation for nucleus--nucleus collisions is much more complicated. There
are no analytic expressions  in the Glauber model for heavy--nuclei
elastic scattering amplitudes. The problem stems from a complicated
combinatorics and from the existence of dynamical correlations related to ``loop
diagrams" \cite {Andreev,Boreskov1}. Thus, usually optical--type approximation
 \cite{Formanek,Czyz} and probabilistic models for multiple rescatterings \cite
 {Pajares} are used. For inclusive cross sections in AB--collisions the
result of the Glauber approximation is very simple to formulate due to the
AGK cancellation theorem. It is possible to prove, for an arbitrary number of interactions
of nucleons of both nuclei \cite{Boreskov2}, that all rescatterings cancel in
the same way as for hA--interactions and only the diagrams of Fig.~5 contribute
to the single inclusive spectrum. Thus a natural generalization of eq.~(\ref{eq1})
for inclusive spectra of hadrons produced in the central rapidity region in
nucleus--nucleus interactions
takes place in the Glauber approximation

\be
\label{eq5}
 E\frac{d^3\sigma^a_{AB}(b)}{d^3p}=T_{AB}(b)~E\frac{d^3\sigma^a_{NN}}{d^3p}
\ee
where $T_{AB}(b)=\int d^2s T_A(\vec{s})T_B(\vec{b}-\vec{s})$ .
After integration over $b$ eq.~(\ref{eq5}) reads

\be
 E\frac{d^3\sigma^a_{AB}}{d^3p}=AB~E\frac{d^3\sigma^a_{NN}}{d^3p}
\label{eq6}
\ee

Note that eqs.(\ref{eq5}),(\ref{eq6}) are valid for an arbitrary set of Glauber diagrams
and are not influenced by the problem of summation of ``loop" diagrams
mentioned above. 

The densities of charged particles can be obtained from eqs.(\ref{eq5}),
(\ref{eq6}) by
deviding them by the total inelastic cross section of nucleus--nucleus
interaction. For example 

\be
\frac{dn^{ch}_{AB}(b)}{dy}=\frac{T_{AB}(b)}{\sigma^{in}_{AB}}
\frac{d\sigma^{ch}_{NN}}{dy}
\label{eq7}
\ee
and
\be
\frac{dn^{ch}_{AB}}{dy}=\frac{AB}{\sigma^{in}_{AB}}\frac{d\sigma^{ch}_{NN}}{dy}
\label{eq8}
\ee
In the following we shall use these results to calculate particle
densities in the central rapidity region at energies of RHIC and LHC.

\section{Regge-Gribov theory and hadronic interactions.}

In this section we will briefly review results obtained as an application of the Gribov
theory to hadronic interactions at very high energies. We will need them in
order to determine inclusive particle spectra in NN--interactions at RHIC
and LHC and in order to illustrate the importance of inelastic diffractive
processes for a selfconsistent treatment of high--energy hadronic interactions.

In the traditional Gribov's reggeon approach it is assumed that the Pomeron is
a Regge pole accompanied by the cuts associated to the multi--Pomeron
exchanges in the $t$--channel. These cuts are important to
restore the unitarity of the theory. Reggeon diagrammatic
technique \cite{Gribov4} and the AGK cutting rules \cite{AGK} allow one to
calculate contributions of many--Pomeron exchanges to scattering
amplitudes and relate them to the properties of multiparticle
production.

An important parameter of the theory is the value of the Pomeron intercept
$\alpha_P(0)$. In the simplest model,
where only the single--Pomeron exchange is taken into account the
intercept determined from the analysis of $\sigma^{tot}$ and elastic
scattering data is $\alpha_P(0)\approx1.08$ \cite{Donnachi}. However, this is only an
effective value of the intercept $\alpha^{eff}_P(0)$, which
describes the energy dependence of total hadronic cross sections
in the currently available range of energies,
  -- $\sigma^{(tot)}\sim s^{\alpha^{eff}_P(0)-1}$.
 An extensive phenomenological analysis, which takes into account
multi--Pomeron exchanges, (see e.g. refs. \cite{KaidalovR,CapellaR})
shows that the Pomeron intercept is substantially larger than the
value $\alpha^{eff}_P(0)$ indicated above. With eikonal
type diagrams only one gets $\alpha_P(0)=1.12\div 1.15$ \cite{21r}.
For a more complete set of diagrams, which include interactions between
exchanged Pomerons (related to large mass diffractive production),
 an even larger intercept of $\alpha_P(0)\approx 1.2$ is obtained \cite{Ponomarev}.

In this approach many characteristics of high energy hadronic
interactions are well described \cite{KaidalovR,CapellaR}. Multi--Pomeron
exchanges are very important for understanding many qualitative features of
experimental data. For example the fast increase of inclusive spectra
in the central rapidity region can be reproduced only if
multi--chain configurations, which are due to cutting of multi--Pomeron
diagrams, are taken into account. For pure pole model the density of
produced hadrons in the central region would be energy independent at
large s. Experimentally in pp--interactions it increases with
energy approximately as $s^{\delta}$ with $\delta\approx 0.13$.

Another manifestation of multipomeron exchanges is an existence of
important long-range correlations (for example forward--backward
correlations). They are firmly established experimentally at high
energies. Long range correlations are closely related to broad
multiplicity distributions. The models based on the reggeon diagrams
technique and AGK--cutting rules give a good quantitative descriptions
of multiplicity distributions including violation of the KNO--scaling
at very high energies.

Let us consider now diffractive production of hadrons at very high
energies. Description of these processes in the Regge--model can be found
in reviews \cite{KaidalovPR,Alberi,GoulianosPR}. The differential cross section
for inclusive single diffraction dissociation can be written in the form
\cite{KaidalovPR}

\be
\label{1}
s_2\frac{d^2\sigma}{ds_2dt}=\frac{(g_{pp}^P(t))^2}{16\pi}
\left|G_P(\xi',t)\right|^2\sigma^{(tot)}_{Pp}(s_2,t)
\ee

\noindent where~ $s_2=M^2$,~ $M$~ is an invariant~mass~ of~ the~ produced~
state,~ $G_P(\xi',t)=\eta(\alpha_P(t)) \exp[(\alpha_P(t)-1)\xi]$,
 $\xi=\ln(s/s_2)$ and $\eta(\alpha_P(t))$ is the
signature factor. The quantity $\sigma^{tot}_{Pp}(s_2,t)$ is the
total cross section for Pomeron--particle interaction. This cross section 
has asymptotic Regge behavior for large mass values of $s_2$ - the squared of diffractively produced
state.
In this case, the cross section for diffraction dissociation is
described by the triple--Regge diagrams (Fig.~6) and has the form

\begin{eqnarray}
\label{2}
&&s_2\frac{d^2\sigma}{ds_2dt}=\frac{(g_{pp}^P(t))^2}{16\pi}
\left|G_P(\xi ,t)\right|^2\sum_kg^k_{pp}(0)r^{\alpha_k}_{PP}
\left(\frac{s_2}{s_0}\right)^{\alpha_k(0)-1}\nonumber \\
 &&E\frac{d^3\sigma}{d^3p}=\sum_kG_k(t)\left ({s_2 \over s} \right )^{\alpha_k(0)-2\alpha_P(t)}
\left(\frac{s}{s_0}\right)^{\alpha_k(0)-1}
\end{eqnarray}

Values of $\sigma^{(tot)}_{Pp}(s_2,t)$ and of the triple reggeon vertices $r^P_{PP}$,
$r^f_{PP}$ have been determined from analyses of diffractive
production in hadronic collisions
\cite{KaidalovR,Alberi,GoulianosPR,26r}.

In the pole approximation with $\Delta=0.08$ the total cross section
of inelastic diffraction increases too fast with energy and strongly
contradicts to experimental data at $\sqrt s\sim 10^3\;GeV$. 
This problem is solved by the inclusion of Regge cuts.
For example in ref.~\cite{Ponomarev} a much slower increase of $\sigma_{D}$,
consistent with recent experimental result, was predicted even for
$\Delta=0.21$ . It is important to note that in Gribov theory the amount
of shadowing is closely related to the magnitude of diffractive processes,
which in their turn are influenced by the shadowing. This  complicated,
 nonlinear problem can be solved only by a systematic and selfconsistent
treatment of both diffractive processes (elastic and inelastic) and
multiparticle production.

\section{Particle densities in heavy--ions collisions at superhigh energies}

Now we will address the question of particle densities in heavy--ions collisions
at energies of future colliders. For central PbPb collisions at LHC ($\sqrt{s} \sim 6$ TeV),
existing Monte-Carlo models (see \cite{Alice}) give a rather broad range of values, of
$dn^{ch}/dy$ at $y^* \sim 0$ --from about 8000 particles for the VENUS model \cite{Venus} to
 $\approx 1400$ particles for the String Fusion Model (SFM) \cite{SFM}. Other 
 Monte--Carlo
 models \cite{Hijing,DPMJ} give predictions within this interval. All these
 models are based on the probabilistic approximation to the Glauber model
 of nuclear interactions and should lead to similar results for inclusive
 particle densities, which according to eqs.(\ref{eq7})-(\ref{eq8}),
  do not depend on details
 of the particular model (for NN-interactions all these models give similar
 predictions, since the extrapolation from the experimentally measured region of energies
 is not large). Below, we shall compare these results of Monte--Carlo models
 and of semi--analytic calculations \cite{Merino}
 with predictions of eqs.(\ref{eq7}),(\ref{eq8}) and shall discuss
  deviations from the Glauber
 approximation build in some of these models. Finally, we will compare them with
 predictions of the complete Gribov theory of nuclear collisions --i.e. including shadowing corrections
due to high mass intermediate states in Fig.~3.
 
 Eq.~(\ref{eq8}) for particle densities integrated over impact parameter
(minimum bias events) can be rewritten as

\be
\frac{dn^{ch}_{AB}}{dy}=n_{AB}\frac{dn^{ch}_{NN}}{dy}
\label{d1}
\ee
where $n_{AB}=\frac{AB\sigma^{in}_{NN}}{\sigma^{in}_{AB}}$. It corresponds
to the average number of collisions in the Glauber model. For $A=B>>1~ n_{AB}$
behaves as $CA^{4/3}$ with $C\approx \frac{\sigma^{in}_{NN}}{4\pi R^2_0}
 ~(R_A=R_0A^{1/3})$. It is well known that eqs.(\ref{eq5}),(\ref{eq6}),(\ref{d1})
  can be applied
to hard processes but in the Glauber approximation they are valid
for soft processes as well. We shall see below that for both soft and hard processes
these equations have to be modified.
 
For characteristics of pp--interaction at $\sqrt{s}=6~TeV$ we take
predictions of the Quark--Gluon String model \cite{KaidalovR} and dual parton model (DPM) 
\cite{KaidalovR,CapellaR} : total inelastic
cross section $\sigma^{in}_{pp}=65~ mb$, inelastic nondiffractive cross section
$\sigma^{in,nonD}_{pp}= 50~ mb$, and for density of charged particles at $y=0$
for nondiffractive interactions $dn^{ch}_{pp}/dy=5.0$. The uncertainty in
these numbers is $\approx 10\%$. The total inelastic cross section for
nucleus--nucleus collisions can be calculated either using a simple geometrical
formula or the optical approximation to Glauber model, which both give at
LHC energies for PbPb collisions a value $\approx 5~barn$ (in these
estimates an increase of the radius of NN--interaction with energy has been
taken into account). Using these numbers and eqs.(\ref{eq8}),(\ref{d1})
we obtain for PbPb collisions at LHC at $y^*=0$ the following numbers
for minimum bias events and central ($b < 3$ fm) collisions, respectively 
\be
dn^{ch}/dy = 2100 \quad , \qquad dn^{ch}/dy = 8500 \quad .
\label{11e}
\ee

These numbers are close to results of the VENUS model \cite{Venus} and to the DPM results of
 ref.~\cite{Merino} (1890 and 7900 respectively, with a slightly different
 definition of central collisions). However, the results of SFM \cite{SFM} without string fusion
 and DPMJET-II \cite{DPMJ} are about twice smaller. For SFM this can be due to the fact
 that this code has limits on the number of produced strings and is not reliable
 at LHC energies \cite{Ferreiro}.

Thus, the Glauber approximation predicts very large densities of charged hadrons
in central heavy--ions collisions at LHC. However, are these predictions
realistic~? In order to answer this question we will consider possible
limitations of the Glauber approximation and also the corrections to the AGK cancellation theorem which are important at high energies.

There are two types of corrections to eqs.(\ref{eq1}),(\ref{eq5}), \\
a) The effects due to energy--momentum conservation \cite{Capella},~--the energy
of the initial hadron is shared by "constituents" (see Fig.~2) and each 
sub--collision happens at smaller energy. These effects are very important in
the fragmentation regions of colliding hadrons (or nuclei) and reduce particle
densities. For $y^*=0$ this reduction decreases as $(1/s)^{1/4}$. It is 
important at SPS energies~; however, at LHC energies in the central rapidity
region this effect is small. It is taken into account in the 
Monte--Carlo models mentioned above.\\
b) Another dynamical effect is important at very high energies when
diffractive production of very heavy hadronic states ($M^2>>m_N^2$) becomes
possible and should be taken into account in the diagrams of Fig.~3. Consider
for example a double rescattering diagram of a proton on a nucleus, which
contains, according to Gribov theory, the diffractive large--mass intermediate
states shown in Fig.~7. It is related to the triple--Pomeron interaction discussed
above and corresponds to an interaction between Pomerons (strings in
the string models of particle production). As the total and inelastic cross
sections of hA and AB--interactions at high energies are close to a black disc
limit due to Glauber--type diagrams, these extra interactions have a small
influence on total cross sections. However, they are very important for
inclusive spectra in the central rapidity region \cite{Kaidalov}, where 
contributions of Glauber rescatterings cancel due to AGK rules.

The diagram shown in Fig.~7 is only one of a large class of diagrams with
interactions between Pomerons. The application of AGK--cutting rules to these
diagrams leads to the diagrams for inclusive cross section in AB--collisions
shown in Fig.~8. Extra shadowing effects related to these diagrams modify
 the A--dependence of the Glauber approximation for inclusive spectra 
 eqs.(\ref{eq1}),(\ref{eq5}) in such a way that the behaviour 
$d\sigma_{AB}/dy\sim AB$ of the Glauber approximation changes to 
$d\sigma_{AB}/dy\sim A^{\alpha}B^{\alpha}$, where $\alpha<1$. For very strong
interaction between Pomerons $\alpha\to 2/3$ . This limit was considered
by O.V. Kancheli a long time ago \cite{Kancheli}. It leads to  universal 
particle densities in pp, pA
and AB--collisions. We will show that due to a rather
week interaction between Pomerons even at LHC energies the value of $\alpha$
is close to 0.9. 

The problem of shadowing for inclusive spectra is not especially related to
soft processes. The same graphs of Fig.~8 are relevant also for hard processes
(production of jets or particles with large $p_T$, heavy quarks, large--mass
lepton pairs, e t.c.). For hard processes, due to the QCD--factorization theorem,
inclusive spectra in nucleus--nucleus collisions are given by convolutions
of hard cross sections with distributions of partons in the colliding nuclei. In
these cases the diagrams of Fig.~8 describe shadowing effects for nuclear
structure functions (i.e. distributions of quarks and gluons in nuclei).
Due to a coherence condition these effects are important only in the region
of very small $x_i$ of partons ( $x_i<< 1/(R_Am_N)$). So these effects are
important only at very high energies, when $x_i\approx M_T/\sqrt{s}$ satisfy
this condition. This condition in terms of $x_i$ of partons coincides with
the condition on  diffractive production of large--mass states discussed above
(see for example \cite{Boreskov3}).

The effects of shadowing are observed experimentally in deep
inelastic scattering on nuclei \cite{28,29}. So in order to test the Gribov
method of calculations of shadowing corrections we will apply it to these
processes (Fig.~9). The first diagram in Fig.~9 corresponds to a
sum of interactions with nucleons of nuclei and is proportional to $A$. The second
diagram (Fig.~9b)) describes the shadowing effect due to a coherent interaction of
a virtual photon with two nucleons of the nucleus and is related to diffractive
production in DIS. This process was measured at HERA \cite{23,24} and was well
described in the model, based on the triple--Regge diagrams of
Fig.~10 \cite{14,25}. The study of diffraction dissociation of a virtual photon at
HERA allows a better determination of triple--Regge couplings compared to
hadronic reactions. This is related to the fact that diffraction dissociation in
DIS is much less influenced by absorptive corrections than diffractive
production in hadronic interactions, where ``effective" vertices are much 
smaller than their ``bare" values (see below, after eq.~(\ref{16e})).

The contribution of a double rescattering term to the
$\sigma_{\gamma^{\ast}p}$ is directly expressed in terms of the
differential cross section for the diffraction dissociation of a
virtual photon in $\gamma^{\ast}N$--interactions:
\be
\label{8}
\sigma^{(2)}=\: -4\pi\:\int\: d^2b\, T^2_A(b)\:\int \:
dM^2\frac{d\sigma^{DD}_{\gamma^{\ast}N}(t=0)}{dM^2dt}F_A(t_{min})
\ee

\noindent where \mbox{$F_A(t_{min})=\exp(R^2_At_{min}/3);~t_{min}\approx-m_N^2x_P^2$}.

Higher order rescatterings are model dependent and in the generalized Schwimmer
model \cite{27} we obtain for the ratio $F_{2A}/F_{2N}$ structure functions $F_2$ nuclei and nucleons, in the
region of small $x$

\be
\label{9}
F_{2A}/F_{2N}=\int d^2b\frac{T(b)}{1+F(x,Q^2)T(b)}
\ee

\noindent with $F(x,Q^2)=4\pi\int dM^2 \left(d\sigma^{DD}_{\gamma^{\ast}N}(t=0)/dM^2dt\right)
\left( F_A(t_{min})/\sigma_{\gamma^{\ast}N}(x,Q^2)\right)$.

Theoretical predictions \cite{Pertermann} for nuclear shadowing based
on the model for 
diffraction dissociation of ref.~\cite{25} are in a good agreement with
NMC--data on nuclear structure functions at very small $x$ \cite{28},
including very accurate data on Sn target  (Fig.~11). The model describes
both the $x$ and $Q^2$--dependence of structure functions of nuclei and allows us
to obtain reliable predictions for structure finctions of nuclei
in the region of $x< 10^{-3}$ (Fig.~12) relevant for nuclear interactions at LHC
energies. Let us mention at this point that while the distributions of
quarks in nuclei are known experimentally and can be rather safely
calculated in the experimentally unmeasured region of $x$ and $Q^2$, the situation
with gluons is much less clear. This problem is related to the distribution
of gluons in the Pomeron, which can be in principle extracted from
experimental data on diffractive production of jets or heavy--quarks in
DIS. However, existing experimental data do not allow a reliable
determination of this contribution. The terms with triple--Pomeron
interaction, shown in Figs.~8,10 lead to a universal shadowing. So in
the following we will assume that the shadowing effects at very small $x$
are the same for quarks and gluons (the same assumptions was made in
refs.~\cite{Hijing,Eskola} while in some papers \cite{Huang,Greiner} it
 is assumed that, due to a larger $ggg$-coupling 
compared to $qqg$-coupling, the shadowing for gluons is larger than for quarks).

  The sum of diagrams of Fig.~5 and Fig.~8 leads to the following expression for
inclusive spectra in nucleus--nucleus interactions.

\be
\label{10}
 E\frac{d^3\sigma^a_{AB}}{d^3p}(b)=\int d^2s f_A(\vec{b})f_B(\vec{s}-\vec{b})
 E\frac{d^3\sigma^a_{NN}}{d^3p}  
\ee
\noindent where, in the Shwimmer model, the function $f_A(b)$ coincides with 
 $\frac{T_A(b)}{1+F(x)T_A(b)}$ introduced above in the calculation of shadowing
 for nuclear structure functions. After integration over impact parameter $b$
 eq.~(\ref{10}) becomes
\be
  E\frac{d^3\sigma^a_{AB}}{d^3p}= F_A(s_A)F_B(s_B)E\frac{d^3\sigma^a_{NN}}{d^3p}
\label{11}
\ee
Where function $F_{A}$ is the same as $F_{2A}/F_{2N}$ defined above and
is equal to $A$ when the triple--Pomeron interactions is switched off.
 At $y^*=0$,  $s_A=s_B=m^a_T\sqrt{s}$ (which corresponds to $x_i=m^a_T/\sqrt{s}$).
 
 Thus, due to interactions
 between Pomerons, the Glauber formula (\ref{11e}) is modified in a simple way
\be
\label{12} 
\frac{dn^{ch}_{AB}}{dy}=\frac{dn^{ch}_{AB,Glaub.}}{dy}\gamma_A\gamma_B
\ee
\noindent where the quantity $\frac{dn^{ch}_{AB,Glaub.}}{dy}$ is the Glauber approximation
result --(eq.~(\ref{d1})) and $\gamma_A$ is the shadowing correction to nuclear
structure function $F_{2A}/AF_{2N}$ shown in Fig.~12. In the following, we will
consider it at a low scale $Q^2\sim 1~GeV^2$ relevant to our problem. For
LHC energies  $\gamma_{Pb}\approx 0.5-0.6$ and the total correction to the
Glauber approximation for PbPb--collisions is $0.25\div 0.36$.
 Formula (\ref{12}) is written for minimum
bias collisions. The result for different values of impact parameter $b$
can be calculated using eq.~(\ref{10}). The calculation shows that for central
collisions this correction differs by less than 10\% from the value indicated
above. Thus there is a suppression by a factor of $\approx 3$ for charged particle
densities at LHC energies compared to the results of the Glauber approximation given in
(\ref{11e}). For PbPb--collisions at RHIC, particle densities are reduced by a factor $\approx 2$
compared to the DPM predictions \cite{Merino}, which do not include shadowing corrections.
However, in this case there are large uncertainties in the calculations of these 
corrections due to a strong dependence of the quantities $\gamma_A$ on $x_i$ in the region of
$x_i\sim 10^{-2}$, relevant for these energies.

Let us emphasize that although our derivation of the shadowing corrections is
based on the study of nuclear structure functions, which is valid for hard
processes, we do not assume that hard processes dominate
particle production at LHC--energies. In fact, the shadowing corrections due
to triple--Pomeron interactions are universal and the same
correction factors in eq.~(\ref{12}) are also obtained for soft interactions 
by a direct calculation of the diagrams of
Fig.~8. In the Shwimmer model we will obtain the same expression 
 $\frac{T_A(b)}{1+F(s,y)T_A(b)}$ for shadowing where the function $F(s,y)$
can again be written as an integral of the ratio of the triple Pomeron cross-section, eq.~(\ref{2}),
over the single Pomeron exchange cross-section $\sigma_P(y)$~: 

 \begin{eqnarray}
F(s,y) &=& \left . 4 \pi \int_{y_{min}}^{y_{max}} dy \ {1 \over \sigma_P(y)} \ {d^2 \sigma^{PPP}
\over dy \ dt} \right |_{t=0} =  \frac{(g_{pp}^P(0))^2 e}{4} \int_{y_{min}}^{y_{max}} d \xi \exp
(\Delta \xi ) = \nonumber \\ && \nonumber \\ &=& \frac{(g_{pp}^P(0))^2 e}{4\Delta} \left [ \exp (\Delta y_{max}) - \exp (\Delta
y_{min}) \right ] \label{16e}
\end{eqnarray}

\noindent where $y_{max} = y_{c.m.}^a + {1 \over 2} \ell n (s/m_N^2)$, $y_{min} = \ell n
(R_Am_N)$. We use the same parameters as in ref.~\cite{25}~: $(g_{pp}^P(0))^2=23~mb$ is the Pomeron--proton coupling, $\Delta\equiv
 \alpha_P(0)-1 = 0.13,~e=r_{PPP}(0)/g_{pp}^P(0)\approx ~0.07$ . Eq. (\ref{16e}) leads then to
practically the same suppression factors as given above. Note that we are using here the same value
of the triple Pomeron coupling as in DIS calculations \cite{25}. As explained there, this value
is about three times larger than the one obtained from a fit of soft hadronic diffraction using
only triple Regge terms (without eikonal unitarization) as in refs. \cite {KaidalovR,Alberi,GoulianosPR}. Such a large
value of the triple Pomeron coupling is required in order to describe diffraction in DIS. In
ref.~\cite{25}, this was justified from the work of ref.~\cite{26r}, where it has been shown that
eikonalization of $\sigma_{SD}$ reduces its value by a factor close to 3 -- implying that the
true (or bare) triple Pomeron coupling is about 3 times larger than the effective one. In DIS,
the eikonal corrections disappear very fast when $Q$ increases and thus the bare coupling is the
relevant one. In single particle inclusive production at high energies, the same is true since
the eikonal corrections are absent due to the AGK cancellation. This is the physical reason for
the large shadowing corrections obtained in this paper.

For collisions of identical nuclei (SS, PbPb) the $A^{4/3}$--dependence of particle densities of
eq.~(\ref{d1}) typical for the Glauber model changes to the behaviour $A^{\delta}$. The  value
of delta is a weak function of energy and
it is equal to $\delta\approx 1.1$ at LHC energies. It means that at these
energies the values of $\alpha$ in the A--dependence of inclusive cross sections
for AB--collisions $d\sigma_{AB}/dy\sim A^{\alpha}B^{\alpha}$ is close to 0.88.
The value of $\alpha$ should slowly decrease as energy increases. In the case
of stronger shadowing for gluons than for quarks a somewhat smaller value of
$\alpha$ can be obtained. 

Let us compare our results with Monte--Carlo calculations, which take into
account shadowing effects \cite{SFM,Hijing}. In the SFM model \cite{SFM} the
interaction between Pomerons is introduced via a mechanism of string fusion
and is estimated from geometrical sizes of strings. The accuracy of such an
estimate is not clear but it leads to a reasonable suppression factor (about 2)
for particle densities at LHC energies (though application of the 
SFM Monte--Carlo is questionable at LHC ,--see above). In the Hijing model
 \cite{Hijing}, existing data on
nuclear shadowing were parameterized and thus the shadowing effects are also
not very diffrent from our predictions, though the Hijing model leads to
a somewhat smaller suppression. This is connected to the choice of a saturation
 for shadowing at small values of $x~(x\sim 10^{-3}) $ in their parameterization.
 The reggeon formalism
 allows us to determine the $x$--dependence of
shadowing at small $x$ and it shows that the shadowing is still increasing
as $x$ decreases even for $x\sim 10^{-4}$. Saturation happens at much smaller
values of $x$.

\section{Conclusions}

Gribov theory of high--energy interactions of hadrons and nuclei is based on
general pro\-per\-ties of amplitudes in relativistic quantum theory and
provides an unified approach to a broad class of processes. In this theory,
the Glauber approximation to nuclear dynamics is valid in the region of not too
high energies and should be modified at energies of RHIC and LHC. AGK--
cutting rules provide a very powerful tool for the study of multiparticle
production for all types of high--energy processes and allow one to obtain
simple predictions for inclusive cross sections in hh, hA and AB--collisions.

In this paper we used AGK--cutting rules to obtain predictions for densities
of particles at future heavy--ions colliders. Gribov theory then allows
to determine corrections to the Glauber approximation for
inclusive particle spectra by relating them to cross sections of large--mass
 diffraction. The technique has been applied to
calculation of shadowing effects for structure functions of nuclei and
a good agreement with experimental data on these processes has been obtained.
The same approach predicts a strong reduction of particle densities at
super--high energies as compared to predictions of the Glauber approximation. Our calculations
show that the DPM results \cite{Merino} for PbPb collisions are reduced by a factor 3 at LHC and 2 at
RHIC energies. The expected values of the charged particle density in central PbPb collisions
at $y^* = 0$ is thus 2500 at LHC and 1000 at RHIC.

Future experiments at RHIC and LHC will test these theoretical 
predictions and will allow a better determination of the parameters that
govern the dynamics of shadowing.
\vspace{0.5cm}

 {\bf Acknowledgments}
 
This work was
supported in part by INTAS grant 93-0079ext., NATO grant OUTR.LG
971390, RFBR grants 96-02-191184a, 96-15-96740 and
98-02-17463 .
\\

\newpage
\bc
{\bf FIGURE CAPTIONS}
\ec

{\bf Fig.~1}  Diagrams of the Glauber model for the elastic $hA$--scattering
amplitude.\\

{\bf Fig.~2} Diagrams for high--energy $hA$--interactions.\\

{\bf Fig.~3} The dispersion representation diagrams for $hA$--elastic scattering
amplitude. A cross on a line means that the particle is on the mass shell.\\

{\bf Fig.~4} The diagram for inclusive cross section of particle $a$ 
in $hA$--collisions.\\

{\bf Fig.~5} The diagram for inclusive cross section in the Glauber approximation
  for AB--collisions.\\
  
{\bf Fig.~6} Triple--regge diagrams for diffractive production of large--mass
states in $pp$--collisions.\\

{\bf Fig.~7} Large--mass diffractive contribution to $pA$--elastic scattering
amplitude.\\

{\bf Fig.~8} Diagrams for inclusive cross sections in AB--collisions, which
take into account interactions between Pomerons.\\

{\bf Fig.~9} Diagrams for $\gamma^{\ast}A$ interactions.\\

{\bf Fig.~10} Triple--regge diagrams for diffractive production in
$\gamma^{\ast}p$--interactions.\\

{\bf Fig.~11} The ratios $\frac{A_1F_{2}^{A_1}}{A_2F_2^{A_2}}$. Experimental
points are from ref.~\cite{28}.\\

{\bf Fig.~12} The ratios $F_2^A/AF_{2N}$ computed from eq.~(\ref{9}). \\
\vspace{1cm}

\newpage
%\centerline{\bf Figure 1}
\vspace{1cm}

\begin{center}
\hspace{-1.2cm}\epsfig{file=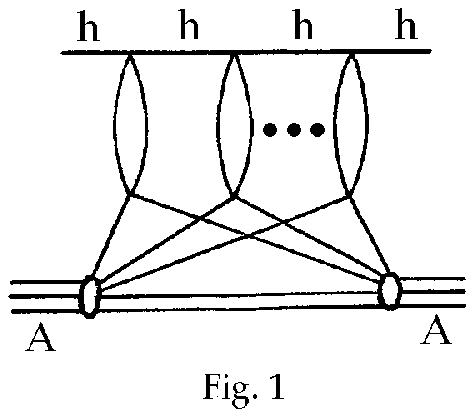,width=13.cm}
\end{center}

%\newpage

%\centerline{\bf Figure 2}
\vspace{1cm}

\begin{center}
\hspace{-1.2cm}\epsfig{file=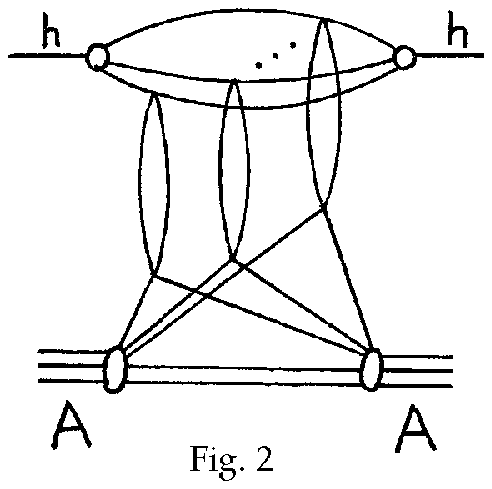,width=11.cm}
\end{center}

%\newpage

%\centerline{\bf Figure 3}
\vspace{1cm}

\begin{center}
%\hspace{-1.2cm}
\epsfig{file=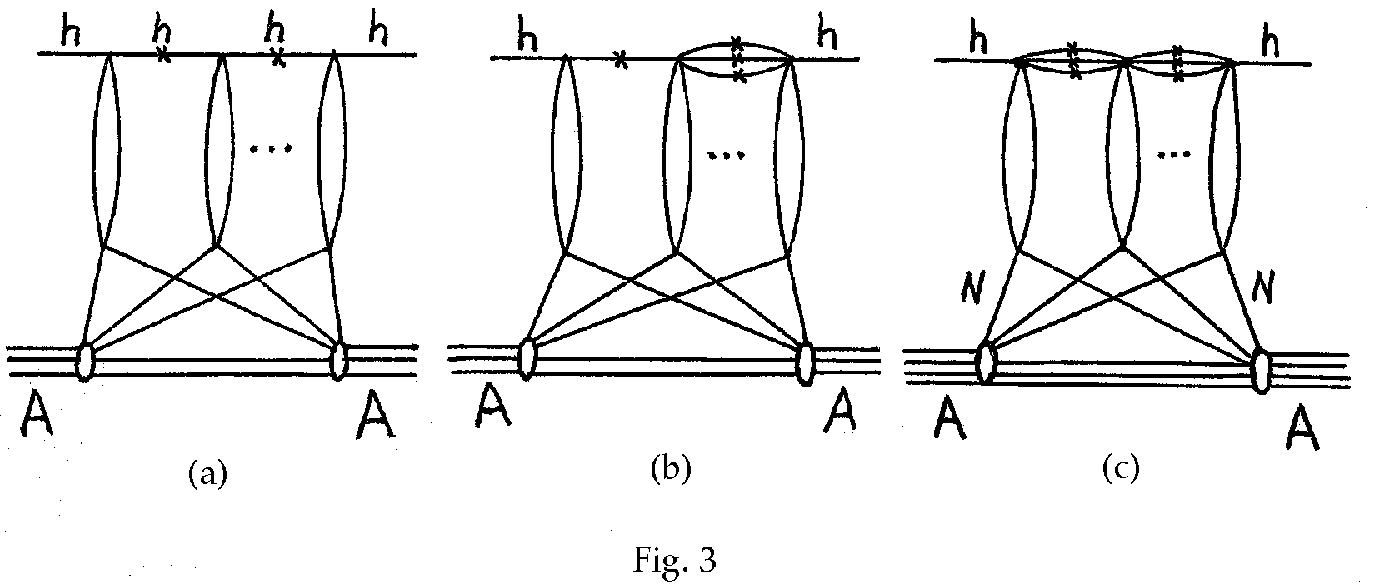,width=13.cm}
\end{center}

%\newpage

%\centerline{\bf Figure 4}
\vspace{1cm}

\begin{center}
%\hspace{-1.2cm}
\epsfig{file=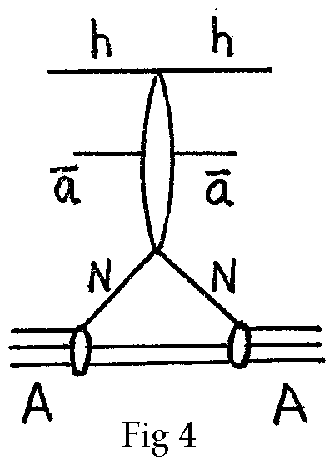,width=7.cm}
\end{center}

%\newpage

%\centerline{\bf Figure 5}
\vspace{1cm}

\begin{center}
%\hspace{-1.2cm}
\epsfig{file=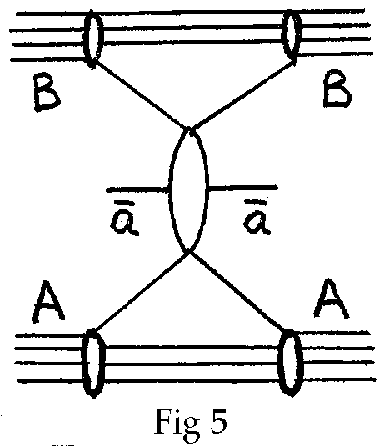,width=7.cm}
\end{center}

%\newpage

%\centerline{\bf Figure 6}
\vspace{1cm}

\begin{center}
%\hspace{-1.2cm}
\epsfig{file=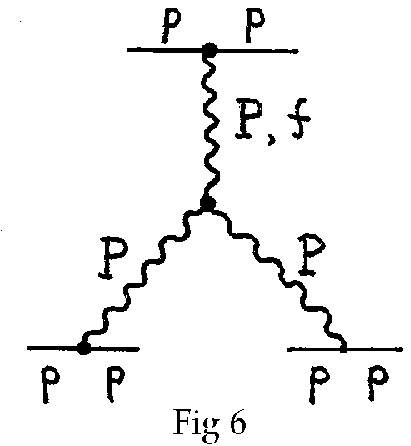,width=7.cm}
\end{center}

%\newpage

%\centerline{\bf Figure 7}
\vspace{1cm}

\begin{center}
%\hspace{-1.2cm}
\epsfig{file=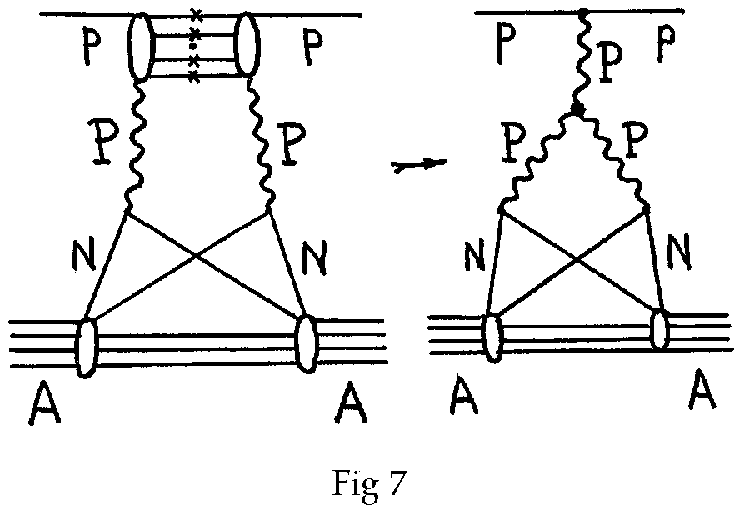,width=13.cm}
\end{center}

%\newpage

%\centerline{\bf Figure 8}
\vspace{1cm}

\begin{center}
%\hspace{-1.2cm}
\epsfig{file=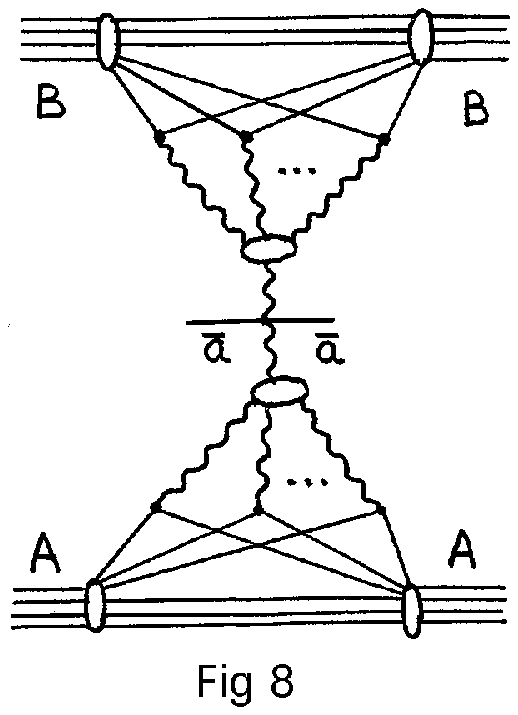,width=7.cm}
\end{center}

%\newpage

%\centerline{\bf Figure 9}
\vspace{1cm}

\begin{center}
%\hspace{-1.2cm}
\epsfig{file=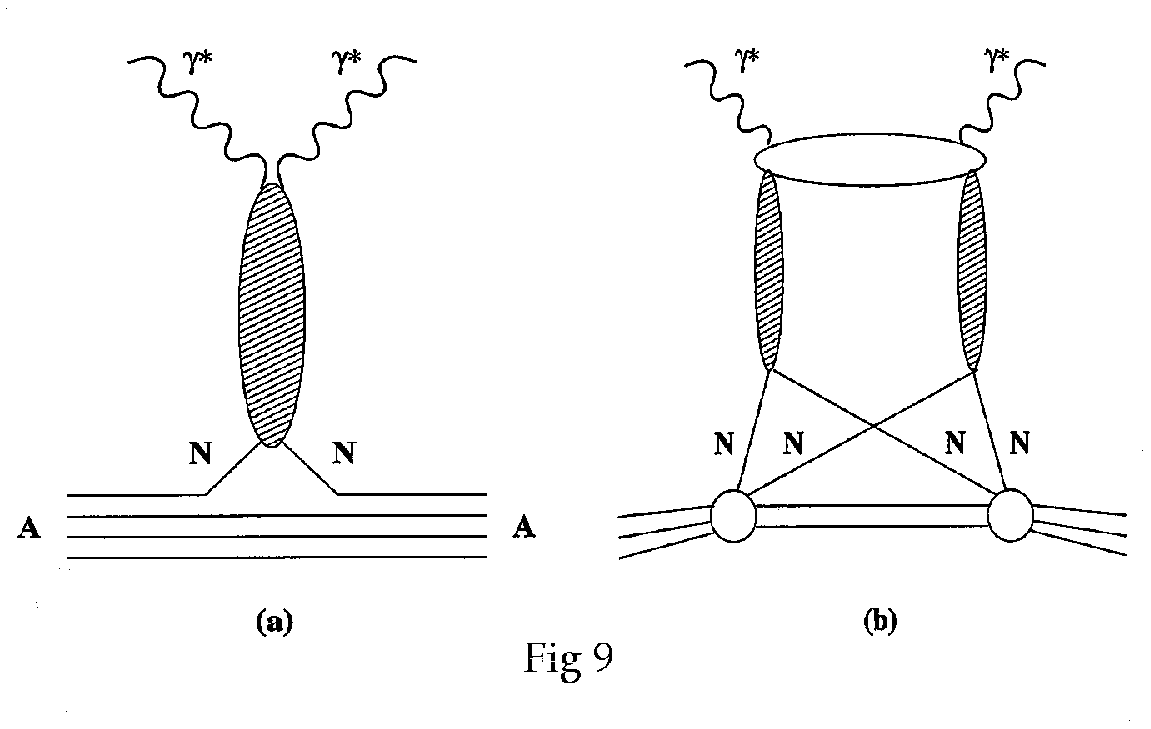,width=13.cm}
\end{center}

%\newpage

%\centerline{\bf Figure 10}
\vspace{1cm}

\begin{center}
%\hspace{-1.2cm}
\epsfig{file=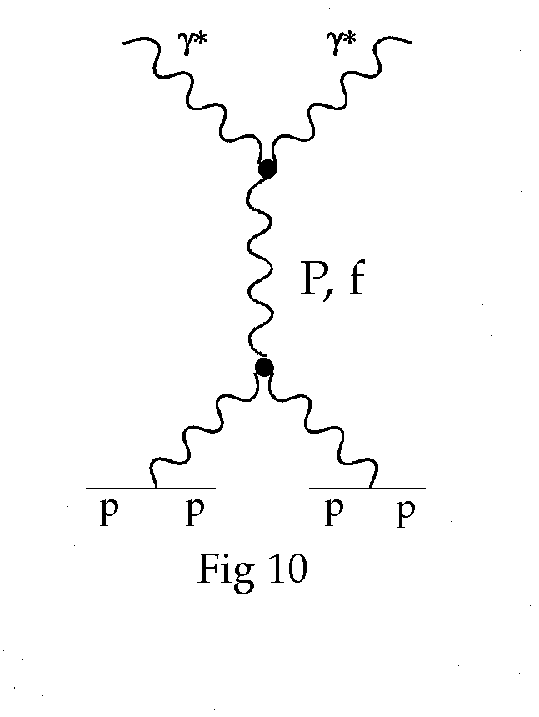,width=7.cm}
\end{center}

%\newpage

\vspace{1cm}

\begin{center}
%\hspace{-1.2cm}
\epsfig{file=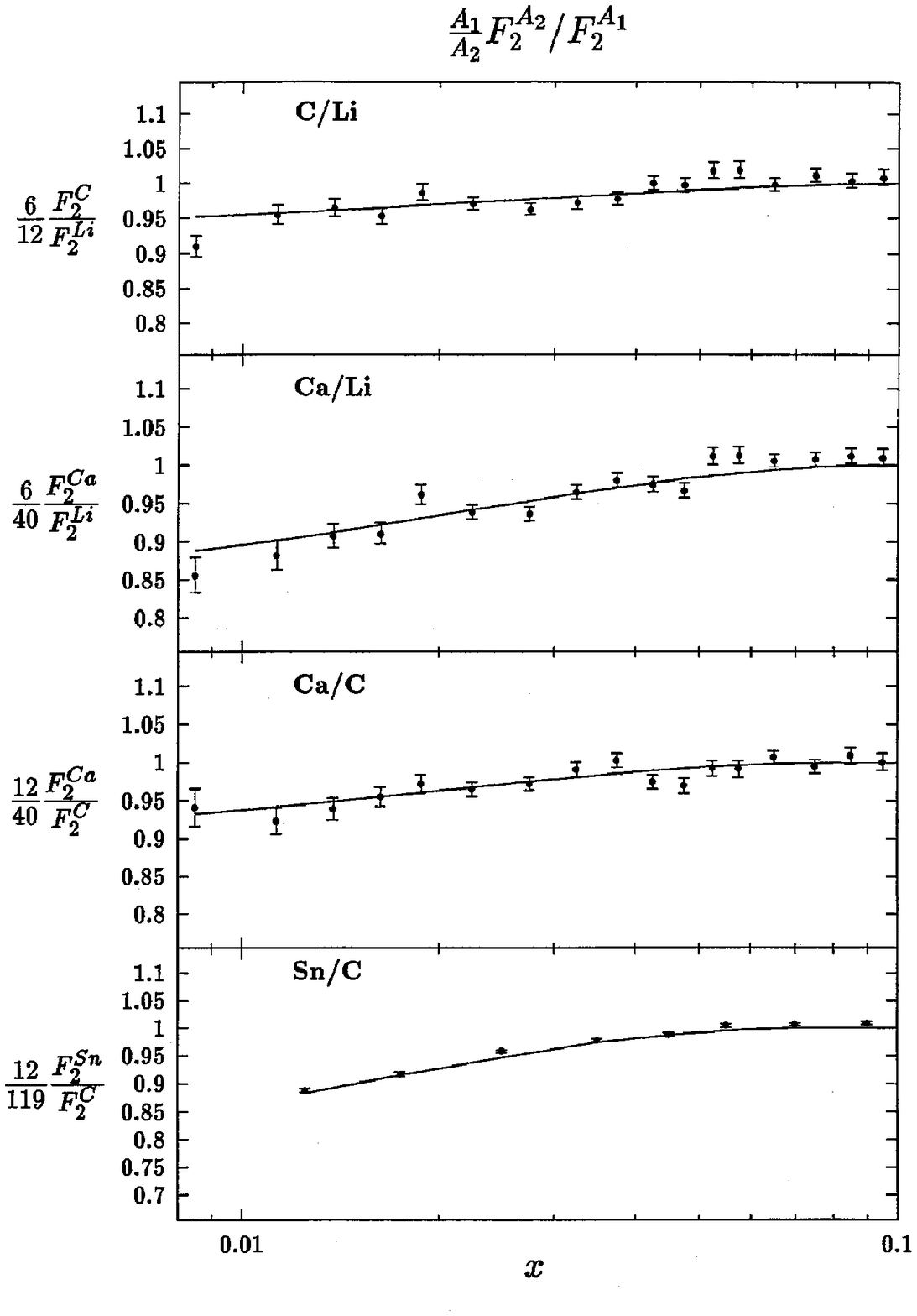,width=10.cm}
\end{center}
\centerline{\Large Fig. 11}

%\newpage

%\centerline{\bf Figure 12}
\vspace{1cm}

\begin{center}
%\hspace{-1.2cm}
\epsfig{file=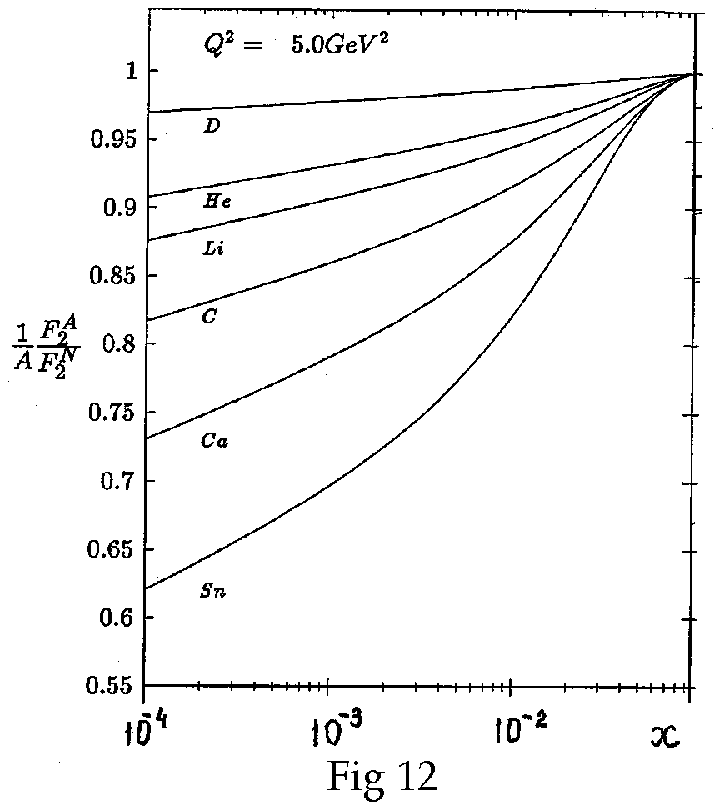,width=10.cm}
\end{center}

\end{document}